\documentclass[twoside,twocolumn,9pt]{article}
\usepackage{extsizes}
\usepackage[super,sort&compress,comma]{natbib} 
\usepackage[version=3]{mhchem}
\usepackage[left=1.5cm, right=1.5cm, top=1.785cm, bottom=2.0cm]{geometry}
\usepackage{balance}
\usepackage{mathptmx}
\usepackage{sectsty}
\usepackage{graphicx} 
\usepackage{lastpage}
\usepackage[format=plain,justification=justified,singlelinecheck=false,font={stretch=1.125,small,sf},labelfont=bf,labelsep=space]{caption}
\usepackage{float}
\usepackage{fancyhdr}
\usepackage{fnpos}
\usepackage[english]{babel}
\addto{\captionsenglish}{%
  
}
\usepackage{array}
\usepackage{droidsans}
\usepackage{charter}
\usepackage[T1]{fontenc}
\usepackage[usenames,dvipsnames]{xcolor}
\usepackage{setspace}
\usepackage[compact]{titlesec}
\usepackage{hyperref}

\usepackage{epstopdf}

\definecolor{cream}{RGB}{222,217,201}

\begin{document}

\pagestyle{fancy}
\thispagestyle{plain}
\fancypagestyle{plain}{
\renewcommand{\headrulewidth}{0pt}
}

\makeFNbottom
\makeatletter
\renewcommand\LARGE{\@setfontsize\LARGE{15pt}{17}}
\renewcommand\Large{\@setfontsize\Large{12pt}{14}}
\renewcommand\large{\@setfontsize\large{10pt}{12}}
\renewcommand\footnotesize{\@setfontsize\footnotesize{7pt}{10}}
\makeatother

\renewcommand{\thefootnote}{\fnsymbol{footnote}}
\renewcommand\footnoterule{\vspace*{1pt}%
\color{cream}\hrule width 3.5in height 0.4pt \color{black}\vspace*{5pt}} 
\setcounter{secnumdepth}{5}

\makeatletter 
\renewcommand\@biblabel[1]{#1}            
\renewcommand\@makefntext[1]%
{\noindent\makebox[0pt][r]{\@thefnmark\,}#1}
\makeatother 
\renewcommand{\figurename}{\small{Fig.}~}
\sectionfont{\sffamily\Large}
\subsectionfont{\normalsize}
\subsubsectionfont{\bf}
\setstretch{1.125} 
\setlength{\skip\footins}{0.8cm}
\setlength{\footnotesep}{0.25cm}
\setlength{\jot}{10pt}
\titlespacing*{\section}{0pt}{4pt}{4pt}
\titlespacing*{\subsection}{0pt}{15pt}{1pt}

\fancyfoot{}
\fancyfoot[LO,RE]{\vspace{-7.1pt}\includegraphics[height=9pt]{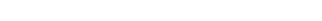}}
\fancyfoot[CO]{\vspace{-7.1pt}\hspace{13.2cm}\includegraphics{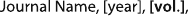}}
\fancyfoot[CE]{\vspace{-7.2pt}\hspace{-14.2cm}\includegraphics{head_foot/RF}}
\fancyfoot[RO]{\footnotesize{\sffamily{1--\pageref{LastPage} ~\textbar  \hspace{2pt}\thepage}}}
\fancyfoot[LE]{\footnotesize{\sffamily{\thepage~\textbar\hspace{3.45cm} 1--\pageref{LastPage}}}}
\fancyhead{}
\renewcommand{\headrulewidth}{0pt} 
\renewcommand{\footrulewidth}{0pt}
\setlength{\arrayrulewidth}{1pt}
\setlength{\columnsep}{6.5mm}
\setlength\bibsep{1pt}

\makeatletter 
\newlength{\figrulesep} 
\setlength{\figrulesep}{0.5\textfloatsep} 

\newcommand{\topfigrule}{\vspace*{-1pt}%
\noindent{\color{cream}\rule[-\figrulesep]{\columnwidth}{1.5pt}} }

\newcommand{\botfigrule}{\vspace*{-2pt}%
\noindent{\color{cream}\rule[\figrulesep]{\columnwidth}{1.5pt}} }

\newcommand{\dblfigrule}{\vspace*{-1pt}%
\noindent{\color{cream}\rule[-\figrulesep]{\textwidth}{1.5pt}} }

\makeatother

\twocolumn[
  \begin{@twocolumnfalse}
{\includegraphics[height=30pt]{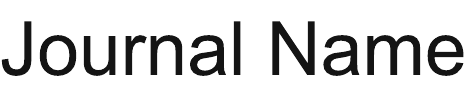}\hfill\raisebox{0pt}[0pt][0pt]{\includegraphics[height=55pt]{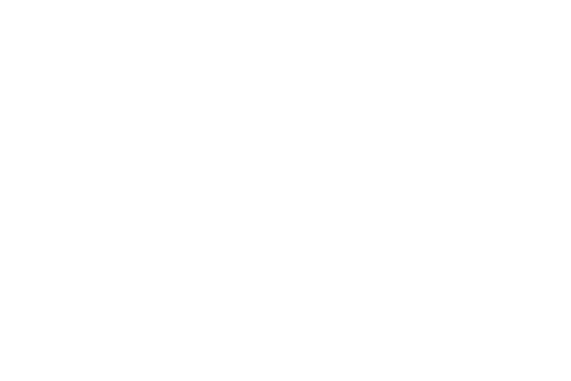}}\\[1ex]
\includegraphics[width=18.5cm]{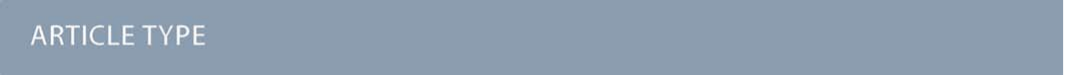}}\par
\vspace{1em}
\sffamily
\begin{tabular}{m{4.5cm} p{13.5cm} }

\includegraphics{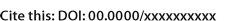} & \noindent\LARGE{\textbf{TopoMAS: Large Language Model Driven Topological Materials Multiagent System$^\dag$}} \\
\vspace{0.3cm} & \vspace{0.3cm} \\

& \noindent\large Baohua Zhang $^{\ast}$\textit{$^{a\ddag}$}, {Xin Li\textit{$^{b\ddag}$},  Huangchao Xu\textit{$^{a}$}, Zhong Jin\textit{$^{a}$}, Quansheng Wu$^{\ast}$\textit{$^{c}$} and Ce Li$^{\ast}$\textit{$^{b}$}} \\

\includegraphics{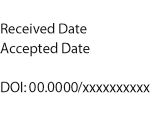} & \noindent\normalsize{Topological materials occupy a frontier in condensed-matter physics thanks to their remarkable electronic and quantum properties, yet their cross-scale design remains bottlenecked by inefficient discovery workflows. Here, we introduce TopoMAS (Topological materials Multi-Agent System), an interactive human–AI framework that seamlessly orchestrates the entire materials‐discovery pipeline: from user-defined queries and multi-source data retrieval, through theoretical inference and crystal-structure generation, to first-principles validation. Crucially, TopoMAS closes the loop by autonomously integrating computational outcomes into a dynamic knowledge graph, enabling continuous knowledge refinement. In collaboration with human experts, it has already guided the identification of novel topological phases SrSbO\textsubscript{3},  confirmed by first-principles calculations. Comprehensive benchmarks demonstrate robust adaptability across base Large Language Model, with the lightweight Qwen2.5-72B model achieving 94.55\% accuracy while consuming only 74.3–78.4\% of tokens required by Qwen3-235B and 83.0\% of DeepSeek-V3’s usage—delivering responses twice as fast as Qwen3-235B. This efficiency establishes TopoMAS as an accelerator for computation-driven discovery pipelines. By harmonizing rational agent orchestration with a self-evolving knowledge graph, our framework not only delivers immediate advances in topological materials but also establishes a transferable, extensible paradigm for materials-science domain.

 } \\

\end{tabular}

 \end{@twocolumnfalse} \vspace{0.6cm}

  ]

\renewcommand*\rmdefault{bch}\normalfont\upshape
\rmfamily
\section*{}
\vspace{-1cm}


\footnotetext{\textit{$^{a}$~Computer Newwork Information Center,Chinese Academy of Sciences, Beijing, China Fax: +86-10-58812114; Tel: +86-10-58812131; E-mail:zhangbh@cnic.cn}}
\footnotetext{\textit{$^{b}$~China University of Mining \& Technology-Beijing, Beijing, China  Email:celi@cumtb.edu.cn}}
\footnotetext{\textit{$^{c}$~The Institute of Physics, Chinese Academy of Sciences,Beijing,China Email:quansheng.wu@iphy.ac.cn}}

\footnotetext{\dag~Supplementary Information available: [details of any supplementary information available should be included here]. See DOI: 00.0000/00000000.}

\footnotetext{\ddag~These authors contributed equally to this work}



\section{Introduction}

    Topological materials, recognized as one of the most significant discoveries in 21st-century condensed matter physics, exhibit extraordinary physical properties such as topologically protected bound states, quantum anomalous hall effect, and quantum transport. These unique characteristics demonstrate tremendous application potential in quantum computing, spintronics, and low-power-consumption devices\cite{hasan2010colloquium,qi2011topological,kitagawa2012observation,chiu2016classification,bradlyn2017topological,bernevig2006quantum,fu2008superconducting}. To date, thousands of topological materials have been computationally predicted or experimentally confirmed\cite{vergniory2019complete,armitage2018weyl,zhang2019catalogue}. This achievement has given rise to a vast family of compounds—including topological semimetals, topological insulators, and topological crystalline insulators. Nevertheless, substantial challenges remain in the efficient discovery and rational design of these materials. Traditional research adheres to a linear workflow—first theoretical prediction, then computational validation. This approach requires exhaustive materials screening and large-scale first-principles calculations, demanding intensive cross-domain collaboration. As a result, information silos and communication gaps frequently hinder progress and can cause promising candidates to be overlooked.

    Generating new crystal structures serves as the foundation for discovering novel topological materials, and recent years have witnessed the emergence of numerous generative models that have brought revolutionary advances to the field of materials design\cite{review}. Early pioneering work includes the Crystal Diffusion Variational Autoencoder (CDVAE) proposed by Xie et al.\cite{cdvae}, which innovatively combined variational autoencoders with diffusion models for periodic material generation. Subsequently, diffusion models have gained widespread application in materials generation, with Jiao et al. developing DiffCSP\cite{diffcsp} for crystal structure prediction through joint equivariant diffusion, and further proposing DiffCSP++\cite{diffcsp++} that incorporates space group constraints to enhance generation quality.

    Transformer-based generative models have also demonstrated remarkable potential. Cao et al. designed CrystalFormer\cite{crystalformer}, which leverages space group information to guide Transformer-based crystalline materials generation, while Chen et al.'s MatterGPT\cite{chen2024mattergpt} achieves multi-property inverse design of solid-state materials. Meanwhile, the application of large language models in materials generation has garnered significant attention. Flam-Shepherd et al. demonstrated that language models can directly generate three-dimensional molecules, materials, and protein binding sites as xyz, cif, and pdb files\cite{cifllm}, and Sriram et al.'s FlowLLM\cite{sriram2024flowllm} combines large language models as base distributions with flow matching.

    Recent research trends indicate that flow-based models exhibit unique advantages in materials generation. Miller et al.'s FlowMM\cite{miller2024flowmm} employs Riemannian flow matching for materials generation, while Luo et al. proposed CrystalFlow\cite{luo2024crystalflow}, a flow-based generative model specifically designed for crystalline materials. Additionally, conditional generation and vector quantization techniques have been widely adopted, such as Con-CDVAE\cite{ye2024cdvae} by Ye et al. and Luo's Cond-CDVAE\cite{cond-cdvae} for conditional crystal structure generation, and VQCrystal\cite{qiu2024vqcrystal} by Qiu et al. that leverages vector quantization for discovering stable crystal structures. 
    
    Advanced training strategies and optimization methodologies have further enhanced the capabilities of generative models. Cao and Wang introduced CrystalFormer-RL\cite{CFRL}, which applies reinforcement fine-tuning to enhance the instruction-following and property-guided generation abilities of transformer-based materials models. Li et al. developed an active learning framework\cite{active1} that combines crystal generation models with foundation atomic models to improve the accuracy and efficiency of inverse design through iterative refinement. Yamazaki et al. presented a multi-property directed generative framework\cite{wyck_gen} that incorporates Wyckoff-position-based data augmentation and transfer learning to handle sparse functional datasets effectively. Ye et al. introduced PODGen\cite{ye2025materialsdiscoveryaccelerationusing}, a highly transferable conditional generation framework that integrates general generative models with multiple property prediction models, demonstrating significant improvements in targeted material discovery efficiency.
    
    Notably, Zeni et al.'s MatterGen\cite{zeni2023mattergen}, published in Nature, provides a general generative model framework for inorganic materials design, while the latest All-atom Diffusion Transformers\cite{joshi2025all} achieves unified generative modeling of molecules and materials, showcasing the cutting-edge developments in this field. Han et al.'s InvDesFlow\cite{active2} for discovering high-temperature superconductors, showcasing the cutting-edge developments and real-world impact of generative models in materials science.

    Recent advances in artificial intelligence (AI) have catalyzed transformative approaches to materials science research, with pre-trained and fine-tuned models increasingly employed for investigating material properties\cite{lee2023towards}. Notable developments include Das's CrysGNN framework—a pre-trained graph neural network that enhances crystalline material property prediction\cite{das2023crysgnn}—and Dong's integrated pipeline that combines transformer-based composition generation, template-guided structure prediction, and GNN relaxation for discovering novel stable 2D materials\cite{dong2023discovery}. In parallel, Huang et al. developed the Materials Informatics Transformer (MatInFormer), which learns crystallographic grammar through space group tokenization\cite{huang2023materials}, while Rubungo et al. introduced LLM-Prop for predicting the physical and electronic properties of crystalline solids from text representations\cite{rubungo2023llm}. Additionally, Song et al. proposed Numerical Reasoning Knowledge Graphs (NR-KG) for structured material property analysis\cite{song2023bridging}. Finally, Xie et al. demonstrated Darwin 1.5—a large language model optimized through two-stage training that processes natural language inputs to accelerate material discovery and property prediction\cite{xie2024darwin}.

    Building upon these AI-driven methodologies, AI agent systems based on large language models (LLMs) have emerged as a transformative paradigm for handling complex tasks and collaborative decision-making in materials science. The Eunomia system by Ansari et al. deploys LLMs as domain-specialized "chemist agents" to extract critical information from complex research literature\cite{ansari2024agent}. Further advancing this approach, Hu et al. established a multi-agent framework for simultaneous optimization of material composition, structure, and properties through symbolic regression\cite{hu2024multi}, complemented by Park et al.'s structure-aware architecture that enables cross-state knowledge retrieval via dynamically allocated specialist agents\cite{park2024leveraging}, and Ghafarollahi's AtomAgents, which integrates LLMs with physics simulators for autonomous alloy design\cite{ghafarollahi2024atomagents}. In practical applications, MatAgent developed by Bazgir et al.\cite{bazgir2025matagent} represents a human-in-the-loop multi-agent LLM framework specifically designed to accelerate the material science discovery cycle, exemplifying the direct application of multi-agent systems in materials research. Ghafarollahi and Buehler further proposed a graph neural network-powered LLM-driven multi-agent system for rapid and automated alloy design\cite{ghafarollahi2024rapid}, expanding the capabilities of physics-aware multi-modal approaches in materials discovery. Ni et al. developed MatPilot\cite{ni2024matpilot}, an LLM-enabled AI materials scientist operating under a human-machine collaboration framework, while their MechAgents\cite{ni2024mechagents} demonstrated that large language model multi-agent collaborations can solve mechanics problems, generate new data, and integrate knowledge across domains. Song et al. proposed a multiagent-driven robotic AI chemist\cite{song2025multiagent} that enables autonomous chemical and materials research on demand, marking a significant breakthrough for multi-agent systems in experimental science. Lai and Pu's Prim\cite{lai2025prim} achieves principle-inspired material discovery through multi-agent collaboration, further validating the value of coordinated AI systems in materials science. Collectively, these efforts confirm that AI agents effectively navigate the inherent complexity and uncertainty of computational materials discovery through intelligent task decomposition and synergistic collaboration.

    Despite significant progress in these pioneering efforts, several critical challenges remain unresolved. Current systems primarily focus on specific tasks, with limited coverage of the comprehensive materials discovery workflow that encompasses materials analysis and reasoning, first-principles calculations, and analytical interpretation. Furthermore, most frameworks exhibit over-reliance on large-scale language models (e.g., GPT-4O\cite{hurst2024gpt}), resulting in prohibitively high computational costs. Reducing system complexity and resource consumption without compromising performance represents a crucial aspect of system design that warrants further investigation.

    In the field of topological materials, the discovery of new topological phases often involves intricate and multi-faceted workflows that require sophisticated coordination between theoretical analysis, computational modeling, and experimental validation. The complex nature of topological material characterization, which encompasses electronic band structure analysis, symmetry considerations, topological invariant calculations, and property predictions, presents unique challenges that current AI systems have not adequately addressed. This work aims to address these challenges by developing an intelligent multi-agent framework that not only assists researchers in improving the efficiency of topological material discovery but also explores the potential of utilizing smaller, more efficient models to tackle complex scientific problems. By investigating how coordinated smaller models can achieve comparable performance to large-scale systems while maintaining cost-effectiveness, we seek to demonstrate that sophisticated scientific tasks can be accomplished through intelligent task decomposition and collaborative reasoning without relying solely on computationally expensive large language models.

    Building upon our previous TopoChat RAG framework\cite{xu2024topochat}, this work introduces Topological materials Multi-Agent System (TopoMAS)—a lightweight multi-agent collaborative framework specifically designed for topological materials research. Our primary contributions include:

\begin{enumerate}
    
    \item \textbf{Hierarchical Multi-Agent Framework:} We propose an  hierarchical multi-agent architecture that leverages intelligent task decomposition and multi-level agent collaboration to attain superior model adaptability. By conducting extensive evaluations across benchmark datasets with diverse complexities, we reveal that smaller models, such as Qwen2.5 - 72B\cite{bai2025qwen2}, can not only match but also surpass the performance of larger LLMs while substantially decreasing computational costs and resource consumption.
    
    \item \textbf{End-to-End Automated Research Workflow:} We establish a comprehensive multi-agent system that assists human scientists in completing complex tasks through natural language interactions, including materials retrieval and reasoning, density functional theory (DFT) calculations, and results analysis. The system supports automatic storage of topological material analysis results in knowledge graphs, enabling knowledge updating and accumulation. This automated workflow has successfully led to the discovery of new topological materials, including SrSbO3.
    
\end{enumerate}
 
    The remainder of this paper is structured as follows: Section 2 presents a detailed overview of TopoMAS's system architecture and key technologies. Section 3 describes the experimental settings and evaluation methods, followed by a discussion of the experimental results and case studies. Finally, Section 4 summarizes the paper and provides insights into future research directions.

\section{Methodology}

\subsection{Overview} 
\begin{figure}
        \centering
        \includegraphics[width=1\linewidth]{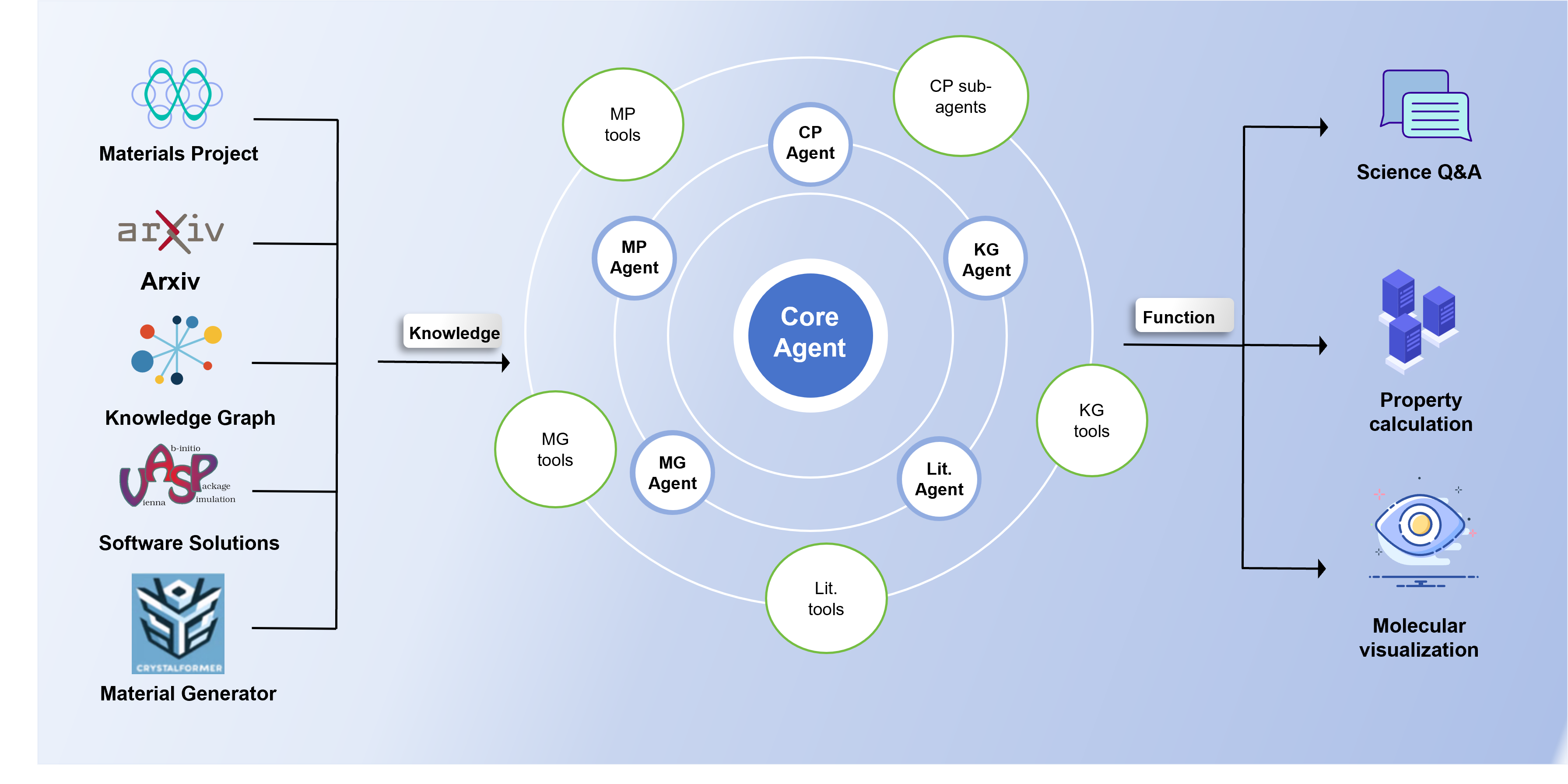}
        \caption{System architecture of TopoMAS, integrating multi-source data resources and hierarchical agent design to automate retrieval, reasoning, and computation tasks in topological materials research }
        \label{fig:1}
    \end{figure}

    TopoMAS employs a hierarchical multi-agent LLM-centric architecture (Fig.\ref{fig:1}) that automates topological materials research through a "perception-reasoning-execution-feedback" closed loop. The architecture is designed based on three core principles: emphasizing both breadth and depth of data coverage, enabling flexible task decomposition and efficient collaboration, and ensuring continuous knowledge accumulation and autonomous evolution. The framework integrates five complementary data repositories spanning standardized material properties (Materials Project), cutting-edge literature (arXiv), our domain-specific TopoKG knowledge graph, first-principles computational software solutions, and crystal structure generation capabilities.

    The TopoMAS system features a three-tiered architecture encompassing central orchestration, specialized execution, and task refinement layers. Serving as the central orchestrator, the \texttt{Core Agent} is responsible for interpreting user intents, decomposing complex tasks, and coordinating agent dispatch. The specialized execution layer comprises five complementary agents— the \texttt{MP Agent}, \texttt{Lit. Agent}, \texttt{KG Agent}, \texttt{MG Agent},
    and \texttt{CP Agent}—that form a cohesive functional ensemble to execute core research functions. At the task refinement level, specialized sub-agents and tools handle detailed computational tasks and validation processes, ensuring comprehensive coverage of the materials discovery workflow.

\subsection{Data Source} 
    To fulfill comprehensive requirements spanning multi-dimensional data retrieval, property computation, and structural visualization in topological materials research, the system integrates five core data resources:

\begin{enumerate}
    \item \textbf{TopoKG Knowledge Graph:} Our curated topological materials knowledge graph integrates heterogeneous data sources encompassing both topological quantum materials and phononic materials, incorporating over 28,000 materials with 37 distinct node types (including basic attributes, topological classifications, photonic properties, and material analogs) and 24 relationship categories\cite{xu2024topochat}. 

    \item \textbf{arXiv Repository:} Enables real-time access to cutting-edge research developments and emerging theoretical insights in materials science.
    \item \textbf{Materials Project Database:} Provides standardized crystal structures and fundamental material properties.
    \item \textbf{Material  Generator:} Utilizes advanced generative models including Conv-CDVAE and CrystalFormer for conditional generation of novel material structures with specified topological properties\cite{ye2024cdvae,crystalformer}.
   
   \item \textbf{Compendium of First-Principles Software Solutions:} A comprehensive collection of VASP\cite{hafner2008ab} computational solutions featuring empirical problem-solving records, parameter optimization strategies, and systematic fixes for common computational challenges encountered in topological materials calculations.
\end{enumerate}

\subsection{Material Retrieval \& Reasoning Subsystem}

    The Material Retrieval \& Reasoning Subsystem orchestrates four specialized agents in a complementary functional architecture:
\begin{figure}
    \centering
    \includegraphics[width=1\linewidth]{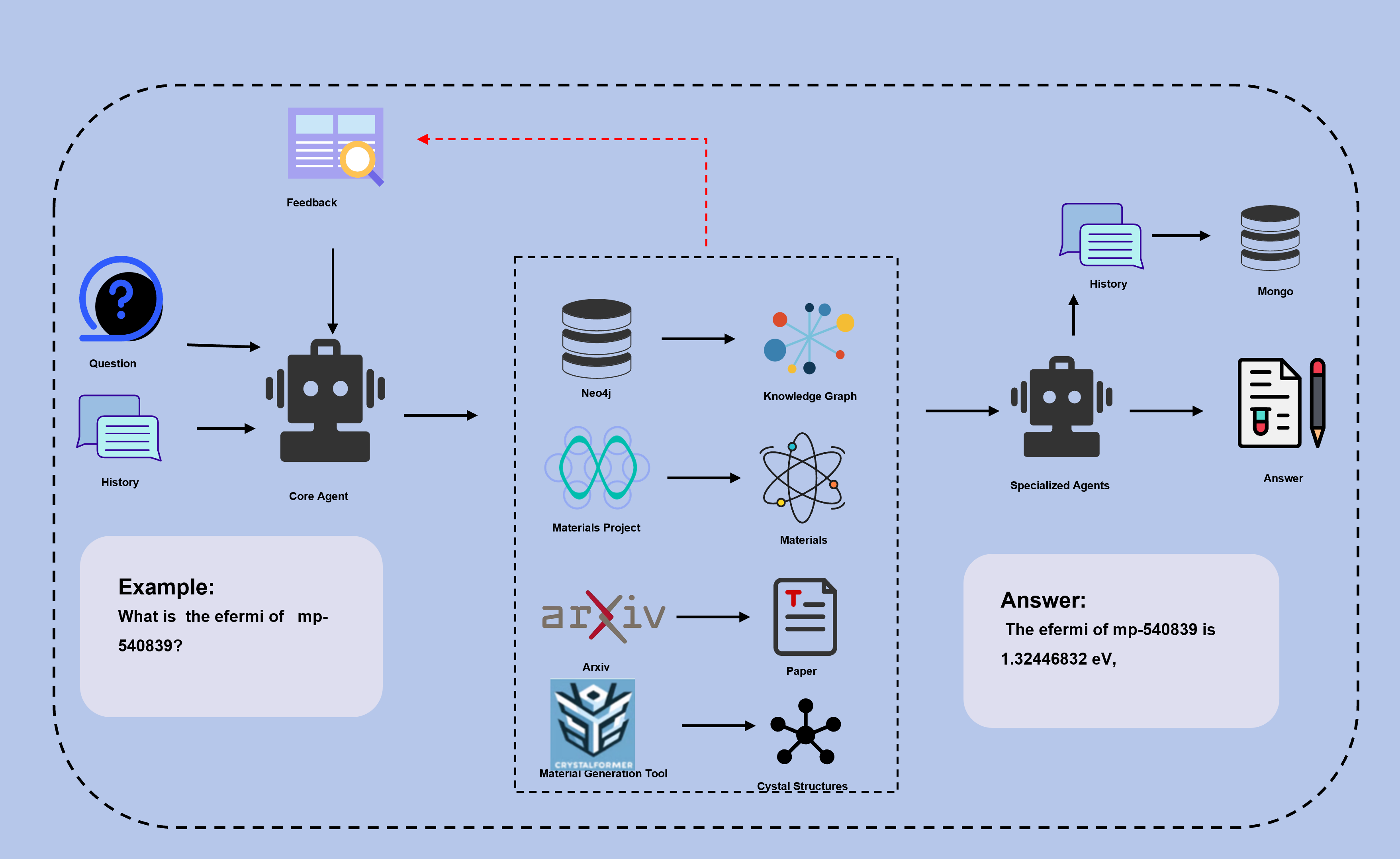}
    \caption{Workflow of material retrieval and \& reasoning subsystem}
    \label{fig:2}
\end{figure}

\begin{itemize}
    \item \texttt{MP Agent}

    This agent executes structured data retrieval and extraction from the Materials Project, one of the world's largest computational materials databases containing over 140,000 entries with extensive API interfaces\cite{jain2013commentary}. However, traditional static query methods prove inadequate for natural language-driven dynamic retrieval. To address this limitation, we designed the ReAct-based \texttt{MP Agent} featuring intelligent query processing through Dynamic Query of Materials Project (DQMP) technology (Fig.\ref{fig:3}).
    
    Our DQMP methodology implements intelligent API encapsulation via a two-stage process:

    1) \textbf{API Documentation Structuring:} Systematically analyze six categories of Materials Project APIs to extract functional descriptions, parameter specifications, and return formats, constructing a structured API knowledge base.

    2) \textbf{Toolchain Automation:} Generate six domain-specific query tools using LangChain's APIChain from the structured documentation:
    
\begin{itemize}
    \item \textbf{Material Summary Tool:} Retrieves basic properties including formula, space group, and lattice parameters
    \item \textbf{Thermodynamics Tool:} Queries thermal properties such as formation energy, phase stability, and chemical potential
    \item \textbf{Elasticity Tool:} Acquires mechanical properties including elastic moduli, Poisson's ratio, and hardness
    \item \textbf{Dielectric Tool:} Extracts dielectric properties such as dielectric constants and optical bandgaps
    \item \textbf{Electronic Structure Tool:} Analyzes electronic structure including energy bands, density of states, and effective mass
    \item \textbf{Magnetism Tool:} Retrieves magnetic properties such as magnetic moments, ordering, and Curie temperatures
\end{itemize}

\begin{figure}
    \centering
    \includegraphics[width=1\linewidth]{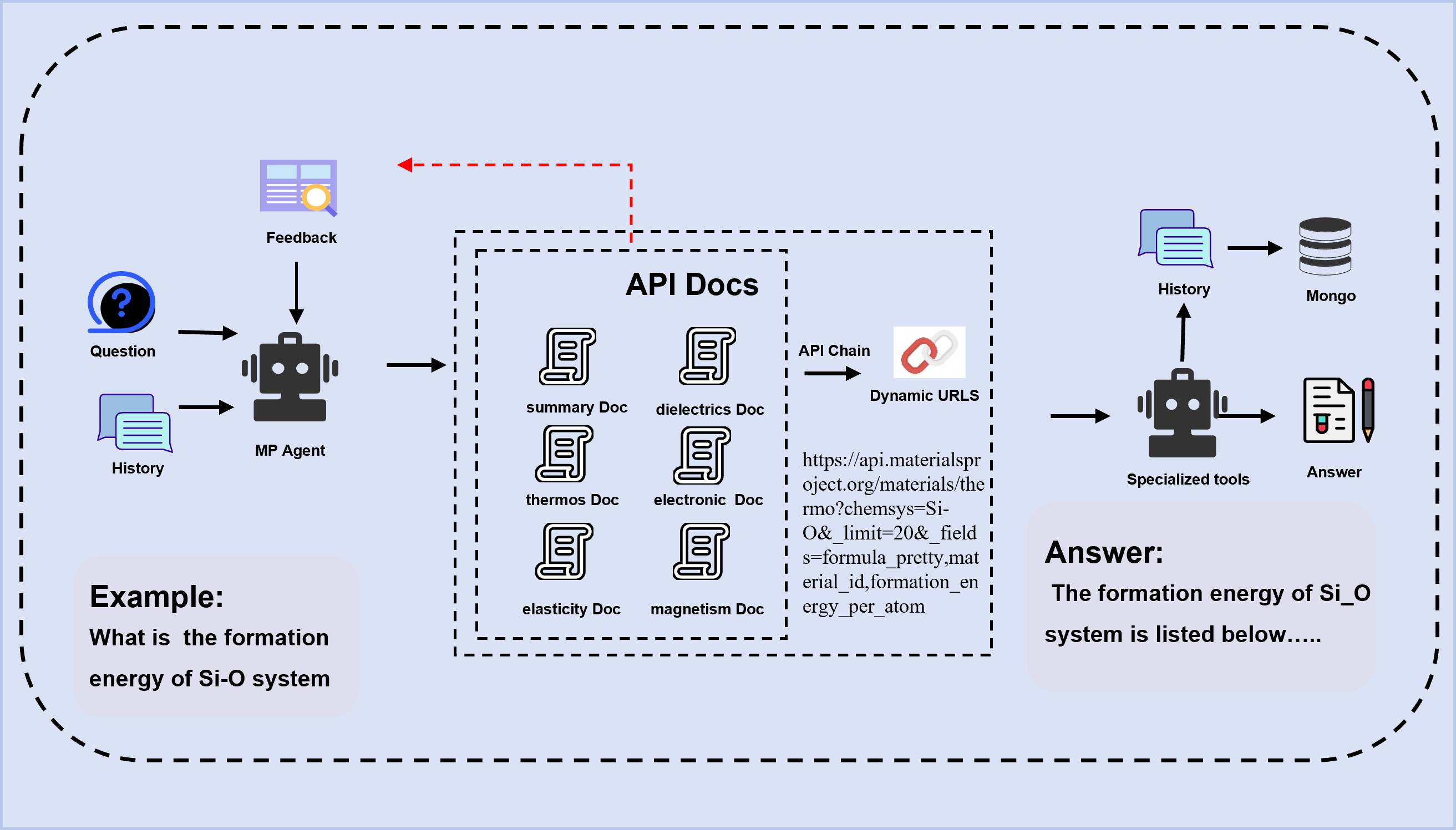}
    \caption{Workflow of dynamic query of Materials Project}
    \label{fig:3}
\end{figure}

This modular design permits seamless integration of additional property-specific tools.

\item \texttt{Lit. Agent} 

    The \texttt{Lit.Agent} designed for semantic search and knowledge extraction from scientific literature in the arXiv repository. It represents an evolution of our prior work\cite{xu2024topochat}, operating on the principles of semantic similarity matching, community detection, and centrality filtering to identify knowledge fragments most relevant to the research query. The agent initially identifies matching documents through semantic similarity analysis. Subsequently, it constructs a similarity network based on abstract embeddings and applies the Leiden community detection algorithm for literature clustering. By selecting core literature with the highest degree centrality from the target community and integrating fine-grained text chunking with similarity ranking, the agent successfully identifies the most valuable knowledge segments for comprehensive literature analysis.
\item \texttt{KG Agent}

    Based on the TopoKG knowledge graph, the \texttt{KG Agent} leverages template-guided LLM optimization to translate natural language topological material queries into Cypher statements, enabling complex searches across material properties and topological features\cite{xu2024topochat}. The agent features a two-stage query optimization process: it first utilizes over 50 pre-encoded description-Cypher pattern pairs for semantic template matching to retrieve relevant Cypher templates, then employs context-augmented prompting based on the selected templates to generate precise Cypher statements for comprehensive knowledge graph traversal.
    
\item \texttt{MG Agent} 
    
   The \texttt{MG Agent} conducts conditional crystal structure generation for designing novel topological materials by utilizing two complementary tools: Conv-CDVAE and CrystalFormer. Conv-CDVAE enables conditional generation of crystal structures with specified properties through a variational autoencoder framework that incorporates compositional and structural constraints\cite{ye2024cdvae}. CrystalFormer leverages space group symmetry information within a transformer architecture to generate crystalline materials while preserving fundamental crystallographic principles\cite{crystalformer}. The agent implements multi-strategy generation capabilities, enabling the systematic creation of materials with user-specified topological properties and structural constraints through these complementary generative approaches.
    
\end{itemize}

The Material Retrieval \& Reasoning Subsystem implements LangChain's ReAct (Reasoning and Acting) framework to address complex inference requirements in topological materials research\cite{mavroudis2024langchain}. This subsystem operates through a structured workflow (Fig.\ref{fig:2}): Initially, the \texttt{Core Agent} interprets user queries through semantic analysis and decomposes complex requests into executable subtasks. Subsequently, the \texttt{Core Agent} dispatches these tasks to appropriate specialized agents based on task taxonomy, whereupon the specialized agents invoke corresponding tools or access relevant data resources to execute specific operations. Finally, the system performs iterative refinement through environmental feedback and evaluation of historical interactions, triggering additional agent invocations until satisfactory results are achieved.

In addition, TopoMAS employs a cascaded ReAct architecture for enhanced reasoning capabilities. The primary ReAct framework manages system-level task decomposition and coordination. Within this hierarchical structure, the \texttt{MP Agent} implements a secondary ReAct framework, functioning as a domain-specific expert (Fig.\ref{fig:3}). This specialized reasoning layer executes intelligent API tool selection, maps abstract queries such as "query material properties" to specific API calls, and prevents upstream agents from experiencing information overload through targeted data filtering.

\subsection{Material Property Computation subsystem} 
The Materials Property Calculation Subsystem constitutes another core component of the TopoMAS framework, addressing complex DFT calculations and topological invariant analysis. To efficiently manage this computational workflow, we developed a multi-agent architecture (Fig.\ref{fig:4}) employing a LangGraph state machine for intelligent, iterative task management. The system ensures reliable execution through four coordinated phases:: 
\begin{figure*}
    \centering
    \includegraphics[width=1\linewidth]{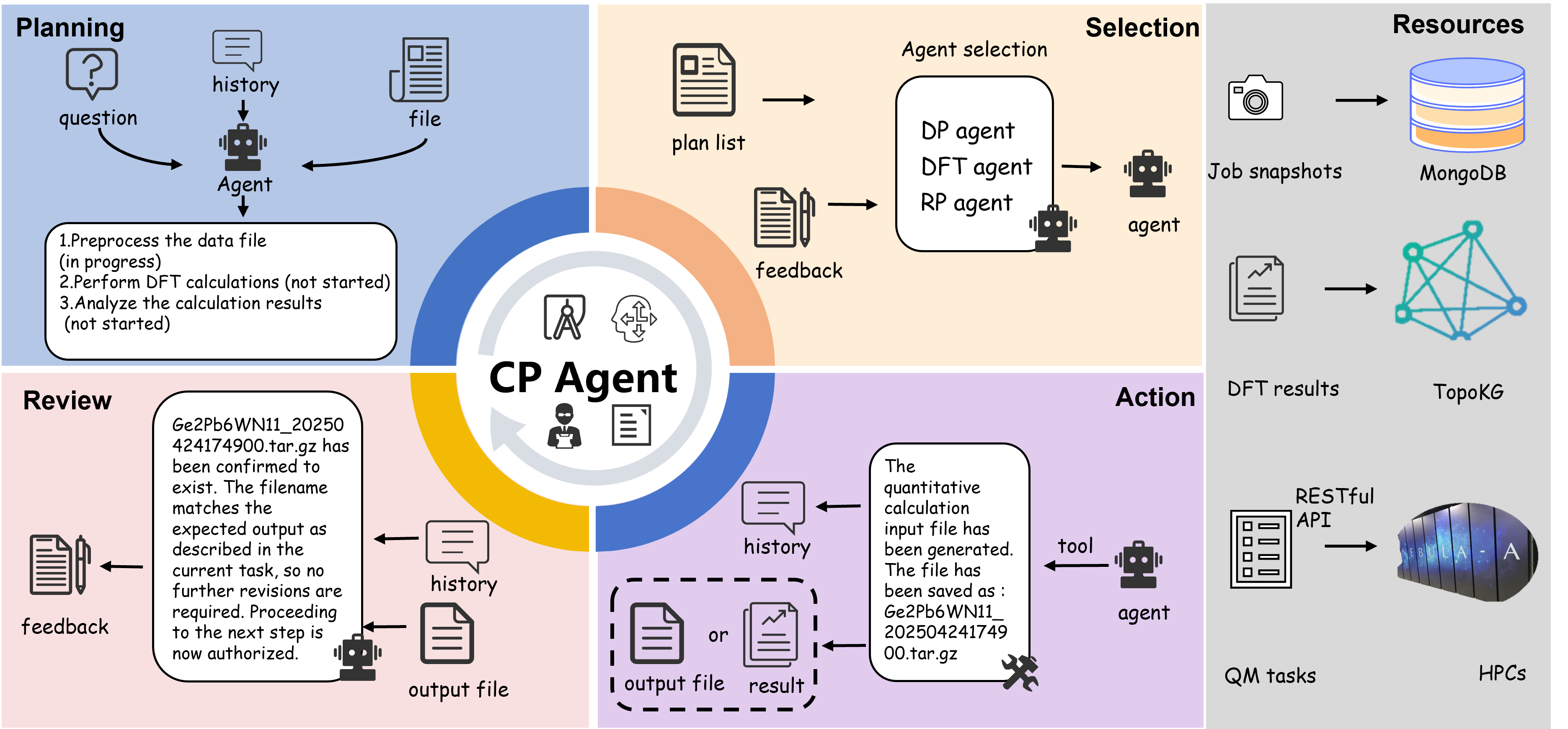}
    \caption{Workflow of material property computation subsystem}
    \label{fig:4}
\end{figure*}

\begin{itemize}

    \item Intelligent Task Planning (Planning)
    
    Serving as the subsystem's central processor, the \texttt{Planning Agent } transforms user requirements into actionable computational workflows. Upon receiving computation requests, it integrates user specifications with historical data and input files, generates optimized task sequences based on computational complexity and available resources, and outputs structured execution plans (plan lists) that specify detailed parameters and resource allocation strategies.

    \item Dynamic Resource Scheduling (Selection)
    
    Agents are selectively activated based on task states and interdependencies. This selection mechanism ensures correct workflow execution by evaluating task types and the completion status of prerequisite tasks.
\begin{itemize}
    \item DP Agent (Data Processing): Parses CIF files and generates input files for target VASP software (e.g., POSCAR/KPOINTS/INCAR) via Symtopo\cite{he2019symtopo})
     \item DFT Agent (DFT Computing): Manages HPC resources through RESTful APIs, enabling asynchronous job submission, queue monitoring, and parallel execution with dynamic load balancing capabilities
     \item RP Agent (Result Processing): Parses computational output files to extract electronic structures and computes topological invariants using SymTopo
\end{itemize}
    
    \item  Task Execution (Action)

    The designated agents execute computational tasks by invoking specialized tools while simultaneously performing real-time state logging to MongoDB for persistent monitoring. This operational framework integrates WebSocket-based progress streaming to provide end users with continuous execution updates. Concurrently, automated error recovery protocols engage through RESTful API monitoring to maintain computational continuity. All computational outputs and processed data are systematically archived in designated storage paths and databases, ensuring immediate accessibility for subsequent analysis and knowledge extraction.

    \item Intelligent Quality Review (Review)

    The \texttt{RE Agent} evaluates execution results through comprehensive quality assessment, examining file integrity to confirm that all necessary output files are generated, verifying computational convergence (e.g., scf-consistent field iteration convergence), and validating result accuracy. If no further revisions are required, the process advances to the subsequent step. When issues are identified, the \texttt{RE Agent} generates targeted improvement recommendations, formulates detailed feedback, and initiates a new Planning-Selection-Action cycle for iterative refinement.
\end{itemize}

The material property computation subsystem employs a circular architecture for iterative optimization, where the canonical workflow progresses sequentially through Planning-Selection-Action-Review phases before advancing to subsequent tasks. When all tasks are completed and validated by the \texttt{RE Agent}, the process terminates. Upon task completion, the \texttt{RP Agent} autonomously extracts structured data, incrementally updating the TopoKG knowledge graph to facilitate continuous knowledge evolution within this self-optimizing material informatics framework (see Fig.\ref{fig:4}). 

To manage extended DFT calculations, a submit-release-poll protocol is implemented alongside nested state management and checkpoint restart capabilities. This framework ensures execution robustness for multi-day ab initio calculations, enabling real-time computational steering through intermediate result analysis. All state transitions are persistently recorded in MongoDB, with automatic snapshot preservation and error logging during exceptions to support scientific reproducibility and agent coordination. System monitoring routines periodically retrieve computational states from MongoDB, delivering progress updates to users via WebSocket streaming.

\subsection{Software and Hardware}
The SymTopo tool is utilized for preparing topological material input files and analyzing topological properties, while the VASP software is employed for DFT calculations\cite{hafner2008ab,he2019symtopo}. The DFT computational processes are scheduled on Era supercomputer via the China National Grid environment\cite{xiao2008sce}.
 
\begin{figure*}
    \centering
    \includegraphics[width=1\linewidth]{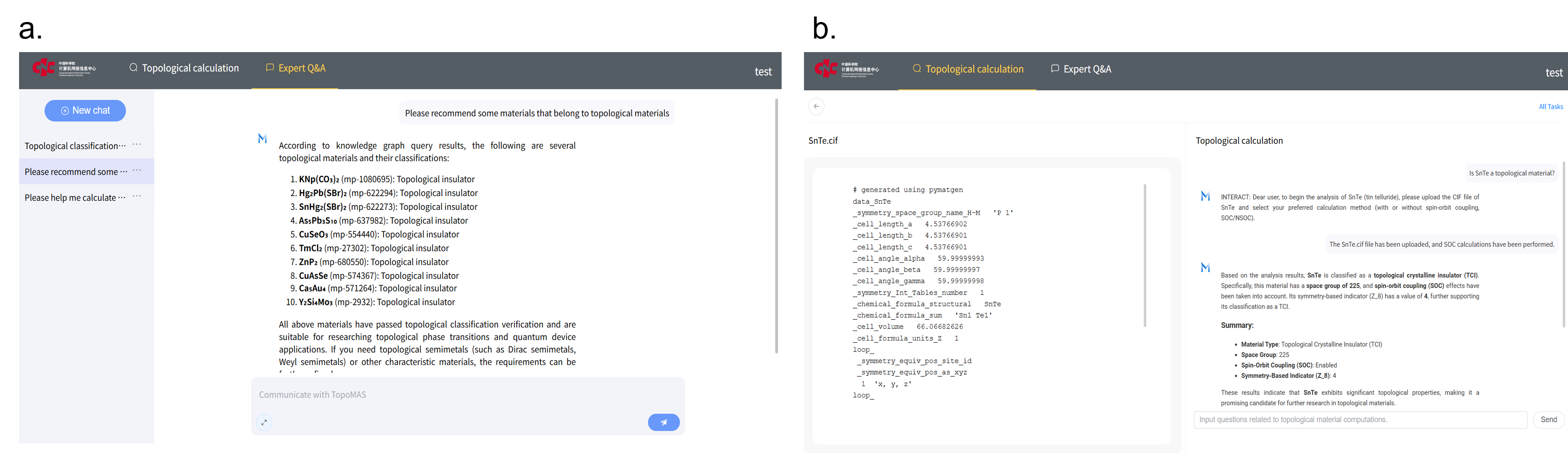}
    \caption{ Web interface of TopoMAS a. Material retrieval \& reasoning  b. Material property computation }
    \label{fig:5}
\end{figure*}

\subsection{Web interface and visualization}

    TopoMAS implements a responsive web interface through a Vue3-based frontend incorporating JSmol for interactive crystal structure visualization, supported by a Flask backend serving RESTful APIs\cite{Robert2013jsmol}. The system utilizes MongoDB for persistent storage of session histories, computational results, and task records, while the TopoKG knowledge graph—continuously enriched with computational outputs—resides in Neo4j to enable advanced semantic retrieval. Real-time synchronization across components is achieved via WebSocket connections, facilitating dynamic content updates and live progress streaming.
    
    To ensure robust concurrency handling during multi-user access, the architecture employs condition variables and event objects for thread coordination, complemented by execution queues that regulate task processing order. The web interface features two integrated modules: (1) A material discovery dashboard enabling natural language queries for multi-dimensional material retrieval alongside structured querying of computational task archives; and (2) A property computation webpage for ab initio computing and topological analysis (see Fig.\ref{fig:5}).
    
    Researchers initiate the topological property computation workflow by submitting a CIF file or providing a Materials Project ID (mp-id), triggering autonomous end-to-end processes. Validated computational outputs are automatically transformed into structured knowledge within TopoKG, creating a self-enhancing repository. For subsequent queries on analyzed materials, this repository enables bypassing repetitive calculations by delivering immediate results directly from the knowledge graph.

\section{Results and Discussion}
\subsection{Experiment details}
\subsubsection{Datasets}
\begin{itemize}
    \item \textbf{LLM4Mat-Bench dataset}: comprises 2,697,779 structure files. Each structure includes three distinct material representations: Composition, CIF, and description\cite{rubungo2025llm4mat}. Notably, 146,143 of these structures originate from the Materials Project. For our testing dataset, we selected 10\% of the test set from the LLM4Mat-Bench paper, specifically entries sourced from the Materials Project. This resulted in approximately 10,318 structures. We performed secondary data cleaning by removing entries with missing attributes and verified the data's accuracy. After rigorous validation, the final test dataset consisted of 8579 entries. These data encompass ten types of properties: formation\_energy\_per\_atom, band\_gap, energy\_per\_atom, energy\_above\_hull, efermi, density, density\_atomic, volume, is\_stable, and is\_gap\_direct. The first eight properties are regression tasks, while the last two are classification tasks. 
    \item \textbf{TopoQA dataset:} 
A question-answering dataset focusing on crystallographic and topological properties of materials. The dataset consists of 110 questions and true answers including space groups, crystal systems, symmetry groups, band gaps, and topological classiffcations.
    \item \textbf{TopoOQ dataset}: 
A open-ended question dataset. It provides 100 open-ended questions, mainly covering topological material recommendations, analysis of material property correlations, similarity-based material recommendations, research trends in topological materials and so on. 

\end{itemize}
\subsubsection{Models and Evaluation Metrics}
Three different LLM models (Qwen2.5-72B\cite{bai2025qwen2}, DeepSeek-V3\cite{liu2024deepseekv3}, and Qwen3-235B\cite{yang2025qwen3}) were assessed across three domain-specific datasets to evaluate their performance within TopoMAS. The evaluation employed domain-specific metrics for comprehensive capability assessment.

For the LLM4Mat-Bench dataset, due to the substantial dataset scale, evaluation was exclusively conducted using the Qwen2.5-72B model with metrics aligned to Choudhary's and Rubungo's benchmark methodology\cite{choudhary2021atomistic,rubungo2025llm4mat}. For the eight regression tasks, performance was measured by the ratio between the mean absolute deviation (MAD) of the ground truth and the mean absolute error (MAE) of the predicted properties, where a higher ratio indicates better performance. We reported the weighted average of MAD:MAE across all properties in each task set (see Equation \ref{eq:wtd_avg_MAD:MAE}).

\begin{equation}
    \text{Wtd. Avg. (MAD:MAE)} = \frac{\sum_{i=1}^{m} \text{TestSize}_i \times \frac{\text{MAD}_i}{\text{MAE}_i}}{\sum_{i=1}^{m} \text{TestSize}_i}
    \label{eq:wtd_avg_MAD:MAE}
\end{equation}
m denotes the number of regression properties in the dataset.

For the two classification tasks, model performance was evaluated using the area under the ROC curve (AUC) for each task and provided the weighted average across all properties(see equation \ref{eq:wtd_avg_auc}).

\begin{equation}
    \text{Wtd. Avg. AUC} = \frac{\sum_{i}^{n} \text{TestSize}_i \times \text{AUC}_i}{\sum_{i}^{n} \text{TestSize}_i}
    \label{eq:wtd_avg_auc}
\end{equation}
n denotes the number of classification properties in the dataset.

For the topoQA dataset, evaluation was conducted across all three foundation models (Qwen2.5-72B, DeepSeek-V3, Qwen3-235B), and the benchmark quantifies system task execution efficiency through four core metrics: (1) Accuracy: determined by expert-validated answer matching against ground truths, measuring fundamental task correctness; (2) Average Time (seconds): derived from mean end-to-end task durations from user input to output generation, reflecting computational efficiency;(3) Median Tokens: monitored via LangSmith to represent typical resource load per interaction; and (4) Total Tokens: instrumented through LangSmith, indicating aggregate system resource consumption. This multidimensional assessment framework enables comparative analysis of model performance across operational quality, speed, and resource utilization efficiency dimensions.

For the topoOQ dataset, the evaluation of the three models (Qwen2.5-72B, DeepSeek-V3, and Qwen3-235B) incorporates efficiency metrics from TopoQA, including Average Time, Median Tokens, and Total Tokens. It also introduces a specialized quality assessment using the third-party DeepSeek R1-0528 evaluator\cite{guo2025deepseekr1}. This objective assessment comprises a three-dimensional framework with 0-10 scoring: (1) Accuracy: semantic alignment with material science principles; (2) Completeness: coverage of multi-aspect query requirements; and (3) Factual Consistency: absence of hallucinated properties. These dimensions collectively contribute to a Composite Score ranging from 0 to 10. This dual-metric approach simultaneously captures operational efficiency and open-domain response quality for complex topological material inquiries.

\subsection{Benchmark on LLM4Mat-Bench dataset}

In the LLM4Mat-Bench evaluation, TopoMAS (based on Qwen2.5-72B) shows better performance in material property retrieval. MatBERT-109M \cite {rubungo2025llm4mat}, Rubungo's top baseline in the MP subset, achieved weighted averages of 5.37 MAD:MAE for regression tasks and  weighted averages of 0.722 AUC for classification tasks (see table\ref{tab:LLM4Mat-Bench}). Our system has made significant progress: regression performance is enhanced by 168.4\% (WTD. Avg. MAD:MAE = 14.421), and classification accuracy is 23.4\% better (WTD. Avg. AUC = 0.8914). Notably, in specific tasks, the MAD:MAE ratios are 23.01 for formation\_energy\_per\_atom, 27.05 for density, and 22.94 for volume, which is over 300\% improvement compared to existing benchmarks. For classification, the AUC reaches 0.92 for is\_stable and 0.86 for is\_gap\_direct, both surpassing the previous standard of 0.722. These improvements are attributed to the \texttt{MP Agent}'s six specialized tools: material\_summary, material\_thermo, material\_elasticity, material\_dielectric, material\_electronic, and material\_magnetism. This focused execution proves that our framework can accurately convert natural language queries into optimized API workflows. Dynamic tool selection algorithms, which sequence operations based on property-type recognition, are the key to this success. This performance paradigm shows that the hierarchical coordination of dedicated agent tools can reach new heights of accuracy in computational materials informatics, effectively connecting semantic understanding with domain-specific execution.

\begin{table}[h]
    \centering
    \caption{The Wtd. Avg. (MAD:MAE) scores for regression task types (the higher the better) and Wtd. Avg. AUC scores for classification task types(the higher the better) for the regression tasks in the LLM4Mat-Bench are reported. Bolded results indicate the best model}
    \begin{tabular}{lcc}
        \hline
        Model & \parbox{2cm}{\centering Regression  (8 task types)} & \parbox{2cm}{\centering Classification (2 task types)} \\
        \hline
        Llama 2-7b-chat:0S & 0.389 & 0.491 \\
        Llama 2-7b-chat:5S & 0.627 & 0.507 \\
        Mistral 7b-Instruct-v0.1:5S & 0.505 & 0.506 \\
        Gemma 2-9b-it:5S & 0.758 & 0.513 \\
        LLM-Prop-35M & 4.394 & 0.691 \\
        MatBERT-109M & 5.317 & 0.722 \\
       \textbf{TopoMAS (Ours)} & \textbf{14.421} & \textbf{0.8914} \\
        \hline
    \end{tabular}
    
    \label{tab:LLM4Mat-Bench}
\end{table}

\subsection{Benchmark on TopoQA dataset }
The TopoQA benchmark evaluation reveals the distinct capability profiles of three foundational models in crystallographic and topological materials analysis (see Table \ref{tab:topoqa_benchmark}). DeepSeek-V3 demonstrates the optimal balance of efficiency and accuracy, achieving the highest accuracy (96.40\%) with the shortest average response time (38.51 seconds). Conversely, Qwen2.5-72B exhibits superior computational economy, maintaining competitive accuracy (94.55\%) while minimizing total token consumption (566,258 tokens). Its median token count (3,754.5) confirms efficient streamlined tool selection, making it particularly effective for routine queries such as bandgap estimation.

Meanwhile, Qwen3's performance is notably less efficient. Despite comparable median token usage (3,710.5 tokens), its tendency toward over-reasoning leads to substantially higher total token consumption (722,308 tokens) and nearly double the latency (79.76 seconds). This inefficiency is particularly evident during combinatorial queries, where redundant tool-call chains cause token overflow, revealing instability in its task-routing mechanisms.

Crucially, Qwen2.5-72B's near-state-of-the-art accuracy (94.55\% vs. 96.40\%) demonstrates that model scale is not the decisive factor in specialized domains. This makes it the optimal solution under strict computational constraints, as it achieves maximal marginal utility where resource efficiency outweighs marginal accuracy gains.

\begin{table}[h]
    \centering
     \caption{Model performance comparison on TopoQA dataset. DeepSeek-V3 demonstrates the best balance of efficiency and accuracy, and Qwen2.5-72B exhibits superior computational economy}
    \begin{tabular}{l c c c c}
        \hline
        Model & \parbox{1cm}{\centering Average Time(s)} & \parbox{1cm}{\centering Median Tokens} & \parbox{1cm}{\centering Total Tokens} & Accuracy \\
        \hline
        Qwen2.5-72b & 39.73 & 3754.5 & \textbf{566258} & 0.9455 \\
        Qwen3 & 79.76 & \textbf{3710.5} & 722308 & 0.9090 \\
        DeepSeek-V3 & \textbf{38.51} & 4040 & 681960 & \textbf{0.9640} \\
        \hline
    \end{tabular}
   
    \label{tab:topoqa_benchmark}
\end{table}

\begin{table*}
\centering
  \caption{Model performance comparison on TopoOQ dataset}
  \label{tab:topooq_benckmark}
  \begin{tabular*}{\textwidth}{@{\extracolsep{\fill}}llllllll}
    \hline
    \textbf{Model} & 
    \parbox{2cm}{\centering \textbf{Average Time}} & 
    \parbox{2cm}{\centering \textbf{Median Token}} & 
    \parbox{1.5cm}{\centering \textbf{Total Token}} & 
    \parbox{1.5cm}{\centering \textbf{Accuracy}} & 
    \parbox{1.5cm}{\centering \textbf{Completeness}} & 
    \parbox{1.5cm}{\centering \textbf{Consistency}} & 
    \parbox{1.5cm}{\centering \textbf{Score}} \\
    \hline
    Qwen2.5-72b & 105.25 & 10467 & 1456993 & \textbf{7.2526} & 7.2421 & 7.9157 & 7.4638 \\
    Qwen3 & 207.06 & \textbf{9492} & 1959332 & 7.2 & \textbf{7.3894} & \textbf{7.9368} & \textbf{7.5078} \\
    DeepSeek-V3 & \textbf{91.34} & 10852 & \textbf{1347943} & 7.1315 & \textbf{7.2000} & 7.9053 & 7.4166 \\
    \hline
  \end{tabular*}
\end{table*}
\subsection{Benchmark on TopoOQ dataset}
The TopoOQ benchmark evaluation demonstrates that the multi-agent framework performs effectively across Qwen2.5-72B, Qwen3, and DeepSeek-V3 models (see Table \ref{tab:topooq_benckmark}). Third-party evaluations confirm consistently high response quality, with scores ranging from 7.2 to 7.9 across accuracy, completeness, and consistency dimensions. This validates the system's robust architecture and toolchain compatibility across different language models.

Performance differences manifest in two key areas. Qwen2.5-72B exhibits slightly superior accuracy (7.25 vs. 7.13--7.20), making it optimal for precise property retrieval. Qwen3, however, excels in complex relational reasoning, achieving unmatched completeness (7.39) and factual consistency (7.94) in open-ended material analysis.

Regarding efficiency, there is a clear trade-off between resource utilization and accuracy. DeepSeek-V3 delivers the fastest performance (91.34s average) with minimal token consumption (1.35M), but this speed comes at the cost of quality compromises across all dimensions. Conversely, Qwen3's intensive verification protocols, while producing comprehensive responses, result in high latency (207.06s) and substantial token overhead (1.96M) due to over-reasoning during multi-agent tool selection.

Crucially, Qwen2.5-72B emerges as the optimal balanced solution, operating at just 1.46× the speed of DeepSeek-V3 while delivering 95\% of Qwen3's analytical depth. This demonstrates that smaller models can achieve high effectiveness (composite score of 7.46) with significantly reduced computational demands. This establishes a practical paradigm for deploying resource-efficient AI systems in specialized domains without compromising core functionality.

Notably, the extended latencies observed in TopoOQ result in average response times ranging from 91 to 207 seconds, from user input to final output. These delays are primarily attributed to several factors: unpredictable delays in third-party API call chains, network instability during arXiv literature retrieval, and the overhead associated with sequential agent coordination.

\subsection{Case study}
\subsubsection{Case 1: Material retrieval \& reasoning workflow}
\begin{figure*}
    \centering
    \includegraphics[width=1\linewidth]{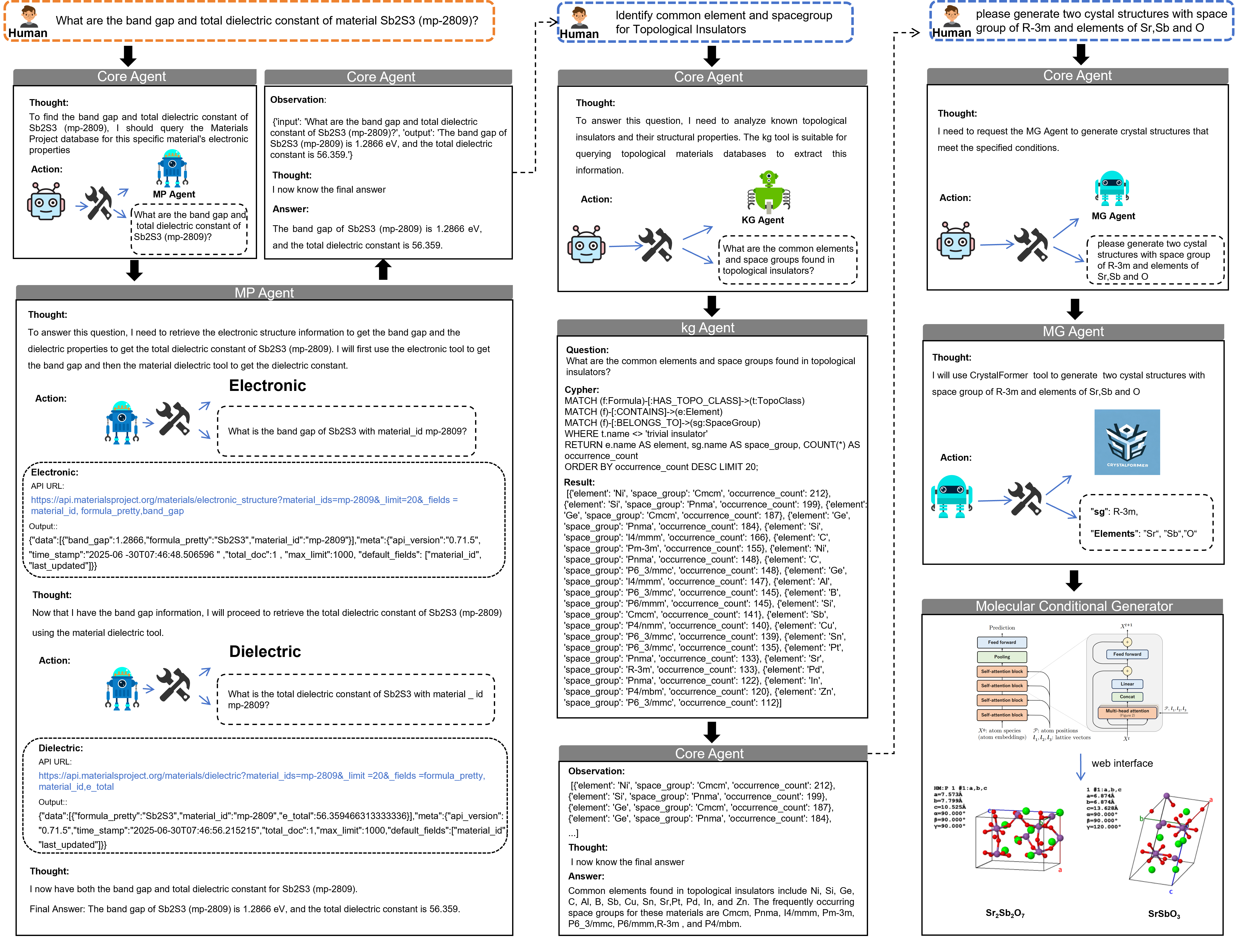}
    \caption{ Material retrieval \& reasoning workflow}
    \label{fig:6}
\end{figure*}

This case illustrates TopoMAS’s  capability through sequential tasks: material basic property query →topological properties analysis → conditional crystal generation(see Fig. \ref{fig:6}).
\begin{itemize}
  
    \item Stage 1: Material basic property query

    Upon receiving the query about the band gap and total dielectric constant of Sb\textsubscript{2}S\textsubscript{3}, the \texttt{Core Agent} directed it to the \texttt{MP Agent}. Using its ReAct framework, the \texttt{MP Agent} executed precise tool queries:
The Material Electronic tool determined the band gap to be 1.2866 eV.
The Dielectric tool calculated the total dielectric constant to be 56.359.
Finally, the \texttt{Core Agent} consolidated and validated these results, producing a concise response.
    
    \item Stage 2: Topological properties analysis
   
    When queried "Identify common element and spacegroup for Topological Insulators", the \texttt{Core Agent} initiates processing by analyzing both the question's semantic intent and functional requirements. Recognizing that the query necessitates topological classification knowledge, it delegates the task to the \texttt{KG Agent}. The \texttt{KG Agent} then performs semantic matching against its Cypher template library, retrieving the 5 most relevant query templates. These templates are incorporated into a structured prompt that directs the LLM to generate an appropriate Cypher query. Upon execution against the materials knowledge graph, the resulting JSON-formatted topological classification data is returned to the \texttt{Core Agent}, which synthesizes and refines the response for final user output.

    \item Stage 3: New crystal structure generation
    
    In response to the request for novel materials with space group R-3m and elements Sr, Sb, and O, the \texttt{Core Agent} identified the parametric design requirements and engaged the \texttt{MG Agent}. The \texttt{MG Agent} utilized its CrystalFormer tool to configure input constraints, processing dual conditions: space group R-3m and elements Sr, Sb, and O. This parametric optimization resulted in two novel crystalline compounds: Sr\textsubscript{2}Sb\textsubscript{2}O\textsubscript{7} and SrSbO\textsubscript{3}. The system then rendered detailed 3D crystallographic visualizations via JSMol in the web interface, providing a comprehensive view of the newly designed materials.

\end{itemize}

\begin{figure*}
    \centering
    \includegraphics[width=1\linewidth]{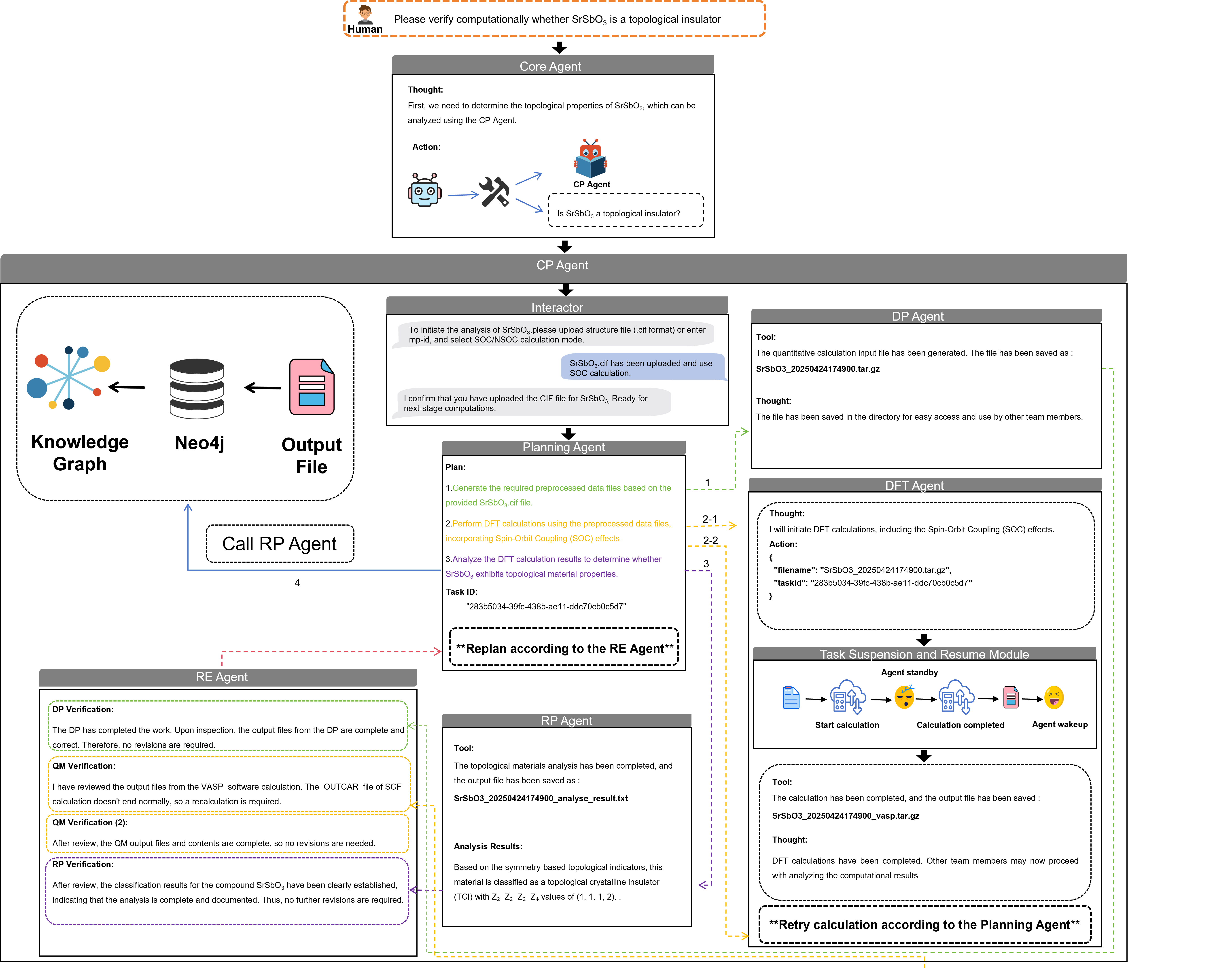}
    \caption{ Topological Property calculation workflow}
    \label{fig:7}
\end{figure*}

\subsubsection{Case 2: Topological Property calculation}
In this case, we performed computational verification for the topological classification of the SrSbO\textsubscript{3} crystal structure generated in the previous case. Upon receiving the request, the \texttt{Core Agent} delegated the task to the \texttt{CP Agent}. The \texttt{CP Agent} initiated its Interactor module to confirm calculation parameters and input requirements with the user. Subsequently, the \texttt{Planning Agent} generated a three-stage execution plan: input preparation, DFT calculation, and topological material analysis. This workflow, implemented via LangGraph's state machine architecture, incorporated robust fault tolerance protocols. The framework coordinated specialized agents across each stage, with persistent state tracking and automated error recovery throughout the process (see Fig. \ref{fig:7}).

\begin{itemize}
    \item Stage 1: Computational Input Preparation
    
    The \texttt{DP Agent} initiates the workflow by processing the crystal structure of SrSbO\textsubscript{3}, converting it into VASP-compatible input files through specialized SymTopo tool integration. This generates the essential calculation files: POSCAR, INCAR and KPOINTS files. Upon file generation, the \texttt{RE Agent} performs comprehensive integrity validation. Following successful verification, the validated input package transitions to stage 2 for DFT calculation execution.
    
    \item Stage 2\_1: DFT calculation execution
    
   The system orchestrates high-performance computing (HPC) execution by submitting the prepared calculation files to the supercomputer via RESTful API. Following submission, the LangGraph workflow engine automatically suspends processing while persisting the current computational state to MongoDB for fault tolerance. Concurrently, the backend monitoring system initiates asynchronous job tracking through periodic API polling to the HPC scheduler, continuously updating task status until the VASP DFT calculation completes. This dual mechanism of state preservation and active monitoring ensures computational continuity while optimizing resource utilization during extended DFT calculations.

    \item Stage 2\_2: Fault recovery protocol demonstration

    To validate the system's resilience, we intentionally modified HPC output files by removing normal termination markers from the OUTCAR file. During validation, the \texttt{RE Agent} detected abnormal computation completion—specifically identifying that the self-consistent field (SCF) calculation failed to terminate properly. This triggered the automated recovery protocol: the \texttt{Planning Agent} initiated workflow rollback to the DFT calculation stage, where the system automatically resubmitted the job. The system persistently iterated this detection-resubmission cycle until the \texttt{RE Agent} confirmed successful output validation, demonstrating robust fault tolerance against computational instability.

    \item Stage 3: Topological verification
    
    The \texttt{RP Agent} conducts essential post-processing on the converged wavefunction outputs: it parses electronic structure data from computational files, calculates topological invariants using the SymTopo algorithm based on DFT calculations, and definitively classifies the material's topological phase. The results confirm that SrSbO\textsubscript{3} is a topological crystalline insulator.

    \item Stage 4: Knowledge integration 
    
    Upon completing Stage 3, the \texttt{Planning Agent} sends the verification results to the web interface through the Interactor module, while initiating backend knowledge consolidation. The \texttt{RP Agent} extracts and structures outputs such as material composition, electronic properties, and topological classification, storing them in Neo4j via the TopoKG ontology framework. This process creates reusable computational records. For subsequent queries, such as whether SrSbO\textsubscript{3} is topological, the \texttt{KG Agent} retrieves pre-computed classifications instantly from the knowledge graph, avoiding redundant calculations and ensuring rapid responses.    
\end{itemize}

\section{Conclusions}
In this work, we introduce TopoMAS, a large language model driven multi-agent framework system for topological materials research. It creates a smart closed-loop framework that integrates multi-source material databases, crystal generators, online resources, and first-principles computation engines, thereby automating the entire "query - reasoning - calculation - update" workflow. Through this framework, we discovered a new topological crystalline insulator SrSbO\textsubscript{3}, enhancing the capability for on-demand design of topological materials. Comprehensive benchmarking across LLM4Mat-Bench, TopoQA, and TopoOQ datasets shows strong model compatibility and task adaptability. The lightweight Qwen2.5-72B architecture achieves over 94\% accuracy on structured tasks and a 7.46/10 composite score for open-ended inquiries. Meanwhile, it consumes 74-83\% fewer tokens than larger models and delivers responses twice as fast. This empirical validation confirms that lightweight models enhanced through multi-agent collaboration and optimized toolchains can transcend traditional computational limitations. They provide cost-efficient solutions for topological materials exploration, effectively bridging computational materials science with cutting-edge artificial intelligence.

 Our future work will involve developing an adaptive computation routing engine and expanding the framework to encompass broader computational domains beyond topological properties. Additionally, we will enhance fault-tolerance mechanisms to facilitate a transition from trial-and-error simulation to demand-driven discovery in materials research.



\section*{Conflicts of interest}
There are no conflicts to declare.



\section*{Acknowledgements}
This work was supported by the National Natural Science Foundation of China (Grant Nos. 22173114 and 22333003) and the Project of Ganjiang Innovation Academy, Chinese Academy of Sciences (Grant No. E455F001).



\balance


\bibliography{rsc} 
\bibliographystyle{rsc} 
\end{document}